\documentclass[
   ,final            
  ]
  {aipproc}
  
  \usepackage{amsmath}
\usepackage{amssymb}
\usepackage{epsfig}
\usepackage{float}
\usepackage{graphicx}
\usepackage{color}

\newcommand{\beq}{\begin{equation}}
\newcommand{\eeq}{\end{equation}}
\newcommand{\beqs}{\begin{eqnarray}}
\newcommand{\eeqs}{\end{eqnarray}}
\newcommand{\beal}{\begin{align}}
\newcommand{\eal}{\end{align}}

\newcommand{\kp}{k_\perp}

\newcommand{\be}{\begin{equation}}
\newcommand{\ee}{\end{equation}}
\newcommand{\bea}{\begin{eqnarray}}
\newcommand{\eea}{\end{eqnarray}}
\newcommand{\bwt}{\begin{widetext}}
\newcommand{\ewt}{\end{widetext}}

\newcommand{\bi}{\begin{itemize}}
\newcommand{\ei}{\end{itemize}}
\newcommand{\ben}{\begin{enumerate}}
\newcommand{\een}{\end{enumerate}}
\newcommand{\bca}{\begin{cases}}
\newcommand{\eca}{\end{cases}}
\newcommand{\bln}{\begin{align}}
\newcommand{\eln}{\end{align}}
\newcommand{\bst}{\begin{split}}
\newcommand{\est}{\end{split}}

\renewcommand{\Im}{\textrm{Im}\,}

\renewcommand\r[1]{(\ref{#1})}

\newcommand\sR{{\ensuremath{{\mathcal R}}}}

\newcommand\sW{{\mathcal W}}
\newcommand\vev[1]{{\ensuremath{\left\langle{#1}\right\rangle}}}

\def\Tr{\mathop{\rm Tr}}

\newcommand\da{{\dagger}}

\def\({\left(}  \def\){\right)}  \def\[{\left[}  \def\]{\right]}  \def\<{\langle}  \def\>{\rangle}

\layoutstyle{8x11single}

\begin{document}

\title{\vspace{-0.28in}Momentum Broadening in Weakly Coupled Quark-Gluon Plasma}


\author{Francesco D'Eramo}{}
\author{Christopher Lee}{}
\author{Mindaugas Lekaveckas}{}
\author{Hong Liu}{}
\author{Krishna Rajagopal}{address={Center For Theoretical Physics, Massachusetts Institute of Technology, MA 02139, USA}}

\begin{abstract}
We calculate the probability distribution $P(k_\perp)$ for the momentum perpendicular to its original direction of motion that an energetic quark or gluon picks up as it propagates through weakly coupled
quark-gluon plasma in thermal equilibrium.
\end{abstract}

\maketitle


As an energetic parton produced in a hard scattering within a heavy ion collision propagates through the strongly coupled matter produced in the collision, it loses energy, it radiates gluons, and it diffuses in momentum-space, picking up momentum $k_\perp$  transverse to its original direction. 
Understanding these processes, and their consequences for jet quenching observables, is important as it has the potential to allow the use of data on these observables
to teach us properties of the medium created during these collisions. In these contributions to the PANIC11 proceedings, as in M. Lekaveckas' talk at the conference, 
we report on our calculation of the probability distribution for $k_\perp$ in the case
where the medium in question is weakly coupled quark-gluon plasma in thermal equilibrium
at temperature $T$.  (The jet quenching parameter $\hat q$, often used to characterize transverse momentum broadening, 
is the second moment of $P(k_\perp)$.)



The general expression for the probability distribution $P(k_\perp)$ that describes transverse momentum broadening of a parton with an energy $E$ that is much greater than
any momentum scales that characterize the medium
has been derived in different ways in 
Refs.~\cite{CasalderreySolana:2007zz,Liang:2008vz,D'Eramo:2010ak}. In the notation of 
Ref.~\cite{D'Eramo:2010ak}, in which the derivation was done using soft collinear effective theory (SCET) that could in the future allow it to be extended beyond the $E\rightarrow\infty$ limit, the result is
\begin{equation} \label{provb}
P(k_\perp) = \int d^2 x_{\perp} \, e^{-i k_{\perp} \cdot x_{\perp}}\,
 \sW_\sR (x_\perp) \quad {\rm with~the~normalization~convention} \quad \int  \frac{d^2 k_\perp}{(2\pi)^2} P(k_\perp) = 1 \ .
\end{equation}
where 
 \beq \label{Wils}
   \sW_{\sR} (x_\perp) \equiv \frac{1}{d\left(\mathcal{R}\right)}
   \vev{\Tr \[ W_{\sR}^\da [x^+ = 0,x_\perp] \, W_{\sR} [x^+ = 0,0] \]}
  \eeq
with
$d\left(\mathcal{R}\right)$  is the dimension of the 
$SU(N)$ representation $\mathcal{R}$  to which the energetic parton belongs, and
$W_{\sR}[x]$ is the Wilson line in the representation $\sR$, namely
\begin{equation}
W_{\sR}\left[y^+, y_{\perp}\right] \equiv P \left\{ \exp \left[i g \int_{0}^{L^-} d y^{-}\, A_\sR^+ (y^+, y^-,y_{\perp})\right] \right\}\ ,
\end{equation}
where $P$ denotes path-ordering and $L = L^-/\sqrt{2}$ is the thickness of the medium through which the parton propagates. The gauge invariance of (\ref{Wils}) can be made manifest by attaching gauge links along the transverse direction from $0$ to $  x_\perp$ at infinity, but these are zero in any covariant gauge, as below. 
Jet broadening has also been analyzed using SCET upon treating the medium via an opacity
expansion, yielding expressions involving only one or two gluon insertions from the 
medium~\cite{Ovanesyan:2011xy}.
The result (\ref{Wils}) sums up arbitrarily many medium gluons, and is valid for any gauge theory medium, whether in equilibrium or not, whether weakly coupled or strongly coupled~\cite{D'Eramo:2010ak}. The nature of the medium enters in the evaluation of (\ref{Wils}), a calculation that has been done for the strongly coupled equilibrium plasma of ${\cal N}=4$ supersymmetric Yang-Mills 
theory previously~\cite{D'Eramo:2010ak}.  Here, we report on the calculation for 
the equilibrium QCD quark-gluon plasma with $T$ so large that physics 
at the scale $T$ is weakly coupled.

\begin{figure}[t]
\includegraphics[scale=0.3]{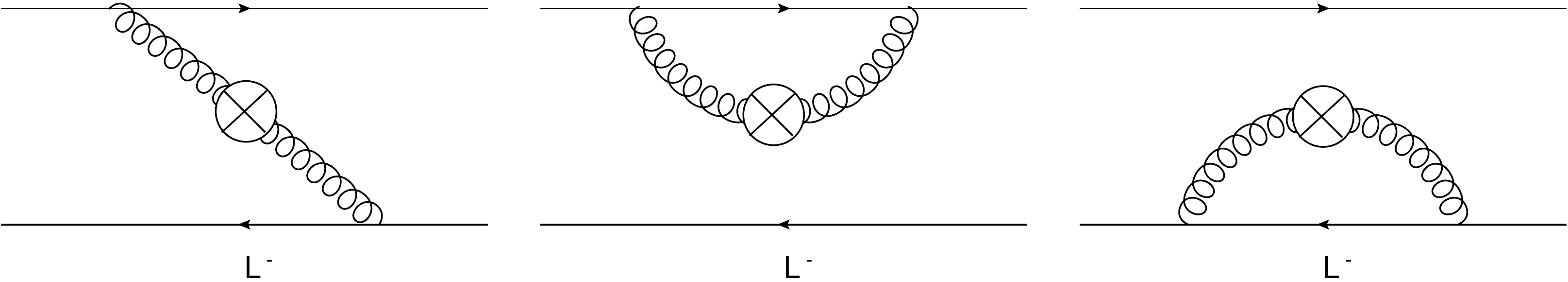}
\caption{Non-trivial contributions to $P(\kp)$, ie $g$-dependent terms in ({\protect\ref{Pcoor1}}). }
\label{Prop}
\end{figure}

Upon assuming weakly coupled plasma, we expand \r{Wils} 
in $g$ and find that the leading terms in \r{provb} are
\begin{align}
P(k_\perp) &= \int d^2 x_\perp e^{-i k_\perp \cdot x_\perp} \Biggl(1 +  \frac{g^2}{d(R)} C(r) \int \int dy_1^- dy_2^-   \text{Tr} \Bigl[ \<A^+ _a(0,y_1^-, x_\perp) A^+_b (0,y_2^-,0)\> - \notag   \\ 
&   \bar{P}\<A^+_a (0,y_1^-, x_\perp) A^+_b(0,y_2^-,x_\perp)\> - P\<A^+_a (0,y_1^-, 0) A^+_b (0,y_2^-, 0)\>  \Bigr] \Biggr)
\label{Pcoor1}
\end{align}
where $\bar{P}$ denotes anti-path ordering.
These contributions to $P(\kp)$ are shown in Fig.~\ref{Prop}. The leading contribution to $P(\kp)$, which we compute, turns out to be ${\cal O}(g^2)$ for $k_\perp < gT$
and  ${\cal O}(g^4)$ for $k_\perp > T$.

Using translation invariance in $x_\perp$ and $x^-$, taking the limit $L \to \infty$, and switching to momentum space, we find
\beq
P(\kp) = \delta^2(k_\perp) g(k_{\perp 0}) \, +\,  \[ P_>(\kp) \]_+^{k_{\perp 0}}, \quad \text{with} \quad P_>(k_\perp) \equiv g^2 C_R L^- \int \frac{dq^-}{2\pi} D_>^{++} (q^+ = 0, q^-, k_\perp)
\label{Pw+}
\eeq
where $g(k_{\perp 0}) \equiv (2\pi)^2 - 2\pi \int_{k_{\perp 0}}^\infty dq_\perp q_\perp P_>(q_\perp)$, and where $[P_>(k_\perp)]_+^{k_{\perp 0}}$ is a 
``plus distribution''~\cite{Ligeti:2008ac}, defined such that it equals $P_>(k_\perp)$ for $k_\perp> 0$, and cancels  the $k_{\perp 0}$ dependence in the delta function term.  This prescription makes $P(k_\perp)$ a manifestly  finite integrable distribution.
%
In (\ref{Pw+}),
$D_>^{++}(0,q^-,k_\perp)$ is the light-cone component of the resummed ``21-propagator", 
related to the retarded gluon propagator by $D_>= 2(1+f(k^0))\, \text{Re}\,D_R$ where $f(x)\equiv 1/(\exp(x/T)-1)$. In Feynman gauge, the retarded gluon self-energy $\Pi^{\mu\nu}_R$  is transverse, $Q_\mu \Pi^{\mu \nu}_R = 0$~\cite{Brandt:1997se}, and
\beq
-iD_R^{\mu\nu}(Q) = \frac{P_T^{\mu\nu}}{Q^2 - G} + \frac{P_L^{\mu\nu}}{Q^2 - F} - \frac{Q^\mu Q^\nu}{Q^4},
\label{resumm}
\eeq
where $Q=(q^+,q^-,q_\perp)$ and we are interested in $q_\perp=k_\perp$, where $P_{T,L}$ are the transverse and longitudinal momentum projectors, and where $F$ and $G$ are the longitudinal and transverse components of $\Pi^{\mu\nu}_R$.
The $++$ component of the tree-level propagator vanishes and for this reason we are interested in the one-loop or, more generally, the resummed propagator \r{resumm}, which does not vanish.
$D_>^{++}$ depends in general on both the real and imaginary parts of the self-energies $F$ and $G$, but on the light-cone ($q^+=0$) and in the UV limit ($q_\perp \gg T$) it depends only on their imaginary parts:
\beq
D_{>,UV}^{++}(0,q^-,q_\perp) = \frac{1+f(q_0)}{q_\perp^2\left( q_\perp^2 + \frac{(q^-)^2}{2}\right) } \(\Im F-\Im G\),
\label{UVDR}
\eeq
meaning that the calculation of $P(k_\perp)$ simplifies at large $k_\perp$.


In the HTL effective theory, which is valid  when all components of $Q$ are ${\cal O}( gT)$ or smaller, $F$ and $G$ are known analytically~\cite{HTL1}.  If any component of $Q$ is larger than 
${\cal O}(gT)$,
we must use the full $F$ and $G$ obtained by calculating the one loop integrals without assuming that internal loop momentum are harder than $Q$.   Note that even if we are evaluating $P(k_\perp)$ at a small value of $k_\perp$ the calculation involves an integral over all values of $q^-$, meaning that the HTL expressions for $F$ and $G$ are not valid throughout the calculation.
In order to obtain an expression for $P_>(k_\perp)$ valid for all $k_\perp$, we evaluate 
$F$ and $G$ without making the HTL approximation, obtaining the imaginary part analytically and the real part numerically. We describe this calculation in a longer paper in preparation.
Using the resummed propagator (\ref{resumm}) in (\ref{Pw+}), we then obtain $P(k_\perp)$ 
at any  $k_\perp$.

\begin{figure}
\includegraphics[scale=0.78]{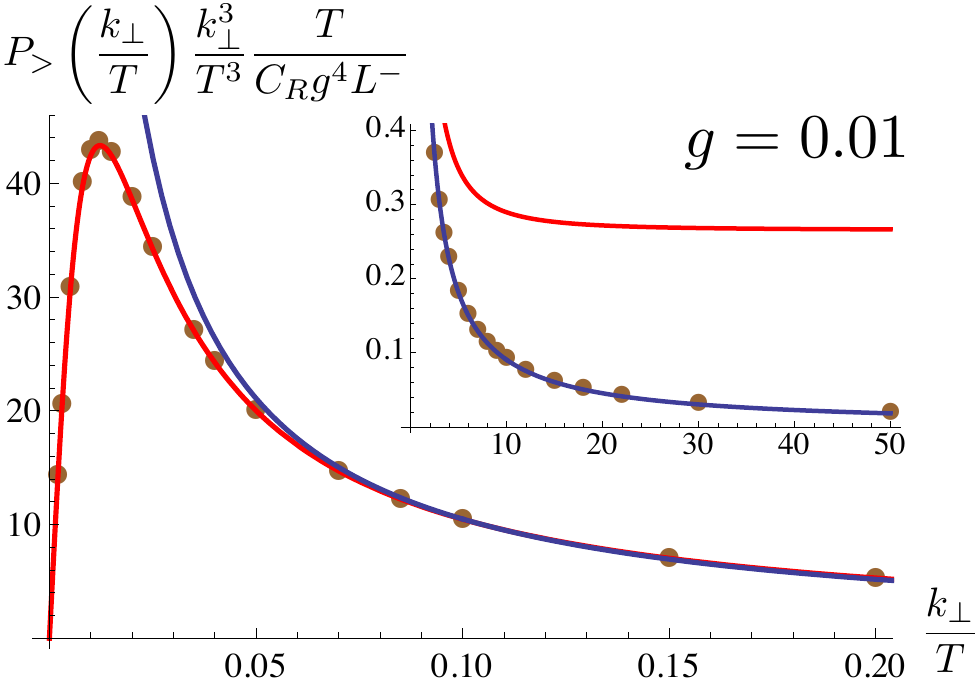}
\includegraphics[scale=0.78]{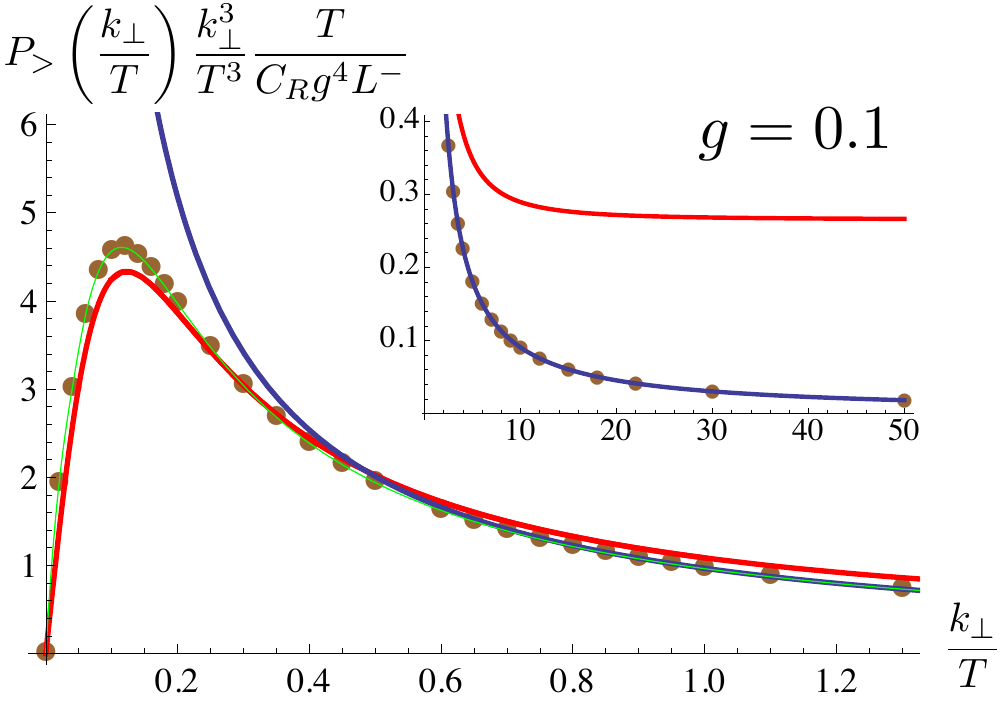}
\caption{
$P_>(k_\perp)$,
as defined in \r{Pw+}, for $g=0.01$ and $g=0.1$. 
We plot $k_\perp^3 P_>(k_\perp/T) $ in units of $g^4 T^2 C_R L^-$.  
The probability distribution for transverse momentum broadening, namely $P(k_\perp)$ defined as a plus function in (\ref{Pw+}), is given by $P_>(k_\perp)$ for $k_\perp>0$ and has a $\delta$-function at $k_\perp=0$ whose weight is such that $P(k_\perp)$ is normalized as in (\ref{provb}). 
The brown dots, joined by a green line to guide the eye,
are the full result obtained by using the resummed propagator \r{resumm}.
The red curve is  obtained by using the HTL approximation for the self-energies $F,G$; at small $k_\perp$, it
agrees with the result in Refs.~\cite{AGZ,CaronHuot:2008ni}. 
The blue curve is the large-$k_\perp$ behavior from \r{UVDR}. For large $k_\perp$, it agrees with the result  in Ref.~\cite{ArnoldDogan}. The insertions in both panels show
the behavior at large $k_\perp$.
}
\label{plots}
\end{figure}

Our results for $P(k_\perp)$ at two values of $g$ are shown in Fig.~\ref{plots}.
In addition to showing our results (the brown dots) we show (as the red lines) the effect of using the HTL approximation for the self-energies $F$ and $G$.  It is no surprise that this approximation is invalid at large $k_\perp$.  It is worth noting that at small $k_\perp$ it comes close to the full result but does not quite agree with it.  We have checked that 
the reason for this is that the HTL self-energies $F$ and $G$ are not correct at large $q^-$ even when $q_\perp$ is small.   The blue line shows the effect of using the propagator \r{UVDR}, valid at large $k_\perp$, with the full 
expressions for \Im $F$ and \Im $G$.  As expected, this approximation breaks down at small $k_\perp$.
The blue curve in Fig.~\ref{plots} is the same in both panels and is in fact $g$-independent.  This confirms that $P(k_\perp)$ is ${\cal O}(g^4)$ at large $k_\perp$.   On the other hand, if we compare our full results (or the red curve, in fact) at the two different values of $g$ in Fig.~\ref{plots} at a value of $k_\perp$ that is small enough that $k_\perp < g T$ in both panels, we see that the quantity plotted is larger by roughly a factor of 100 when $g=0.01$ than when $g=0.1$, indicating that
 $P(k_\perp)$ is ${\cal O}(g^2)$ at such small values of $k_\perp$.

We now compare our results  to those in the literature, first in the small $k_\perp$ limit.  In this regime, the red curve agrees with Refs.~\cite{AGZ,CaronHuot:2008ni}, where HTL self-energies are 
used and where
$f(q_0)$ is approximated as $f(q_0) \sim T/q_0$. Both these approximations
are valid at small $q_0$ but not throughout the integration over $q_0$ that arises in (\ref{Pw+}). We see by comparing the red curve to our results that making  these approximations introduces only a relatively small error.
For $k_\perp \gtrsim T$, 
our results agree with the blue curve in Fig.~\ref{plots} and therefore with Ref.~\cite{ArnoldDogan}. 
So, our results provide 
a small correction to results in the literature 
for $P(k_\perp)$ at small $k_\perp$ and provide a consistent
way of connecting the small $k_\perp$ results to previously known results for $P(k_\perp)$
at large $k_\perp$.  We can also compare our results for the jet quenching parameter 
\beq
\hat q \equiv \frac{1}{L} \int \frac{d^2 k_\perp}{(2\pi)^2} k_\perp^2 P(k_\perp) =  \frac{1}{L} \int_0^\infty \frac{d k_\perp}{2\pi} k_\perp^3 P_>(k_\perp)
\label{qhat}
\eeq
to those in the literature.  (The integrand in (\ref{qhat}) is proportional to the quantity plotted
in Fig.~\ref{plots}.)  Because $P(k_\perp)$ is ${\cal O}(g^2)$ only for $k_\perp < gT$, we find
$\hat q$ of order $g^4$, as is well-known. 
$\hat q$ has been calculated to order $g^4$~\cite{ArnoldXiao} and to 
order $g^5$~\cite{CaronHuot:2008ni}. We find that $\hat q$ in Ref.~\cite{ArnoldXiao} differs
from that we obtain from  $P(k_\perp)$ 
 by only
 0.4\% (4.5\%) at $g=0.01$ ($g=0.1)$. 
 

Note that $k_\perp^3 P(k_\perp)\propto 1/k_\perp$ at large $k_\perp$, meaning that in order to evaluate $\hat q$ (and compare to Ref.~\cite{ArnoldXiao}) we must introduce an ultraviolet cutoff.
The probability of picking up large  $k_\perp$ is significant: 
$P(k_\perp)$ is a correctly normalized probability definition but it has a ``fat tail'', 
such that $\langle k_\perp \rangle$ is finite while $\langle k_\perp^2\rangle\sim \hat q$ diverges.
This behavior  characterizes a weakly coupled plasma that contains
point-like quasiparticles.
We can also compare our weak coupling results to 
$P(k_\perp)$ for a strongly coupled plasma, which turns out to be a Gaussian function of $k_\perp$~\cite{D'Eramo:2010ak}. 
Hence, even if a strongly coupled plasma causes more momentum broadening than a weakly coupled plasma in some average sense, the probability that the hard parton picks up $k_\perp$ is always greater in a weakly coupled plasma than in a strongly coupled one at large enough $k_\perp$. In fact, 
$\hat q$ is greater in a weakly coupled plasma, where it diverges, than in a strongly coupled 
plasma, where it does not.
This reflects the fact that the strongly coupled plasma is a liquid containing no quasiparticles.





\bibliographystyle{aipproc}   
%

\begin{thebibliography}{99}


 


\bibitem{CasalderreySolana:2007zz}
  J.~Casalderrey-Solana and C.~A.~Salgado,
  Acta Phys.\ Polon.\  B {\bf 38}, 3731 (2007)
  [arXiv:0712.3443 [hep-ph]].
  
\bibitem{Liang:2008vz}
  Z.~T.~Liang, X.~N.~Wang and J.~Zhou,
  Phys.\ Rev.\  D {\bf 77}, 125010 (2008)
  [arXiv:0801.0434 [hep-ph]].

  
\bibitem{D'Eramo:2010ak}
  F.~D'Eramo, H.~Liu, K.~Rajagopal,
  Phys.\ Rev.\  {\bf D84}, 065015 (2011).
  [arXiv:1006.1367 [hep-ph]].
 
\bibitem{Ovanesyan:2011xy}
  G.~Ovanesyan and I.~Vitev,
  JHEP {\bf 1106}, 080 (2011)
  [arXiv:1103.1074 [hep-ph]].
  
\bibitem{Ligeti:2008ac}
  Z.~Ligeti, I.~W.~Stewart, F.~J.~Tackmann,
  Phys.\ Rev.\  {\bf D78}, 114014 (2008).
  [arXiv:0807.1926 [hep-ph]].

%
  
\bibitem{Brandt:1997se}
  F.~T.~Brandt and J.~Frenkel,
  Phys.\ Rev.\   {\bf D56}, 2453 (1997)
  [arXiv:hep-th/9703165].
  
  
  
\bibitem{HTL1}
  E.~Braaten, R.~D.~Pisarski,
  Nucl.\ Phys.\  {\bf B339}, 310 (1990); 
  {\it ibid}  {\bf B337}, 569 (1990);
%
  J.~Frenkel, J.~C.~Taylor,
  Nucl.\ Phys.\  {\bf B334}, 199 (1990).

  
  
  

  
\bibitem{AGZ}
  P.~Aurenche, F.~Gelis and H.~Zaraket,
  JHEP {\bf 0205}, 043 (2002)
  [arXiv:hep-ph/0204146].
  
  
  
  
\bibitem{CaronHuot:2008ni}
  S.~Caron-Huot,
  Phys.\ Rev.\  {\bf D79}, 065039 (2009).
  [arXiv:0811.1603 [hep-ph]].
  
\bibitem{ArnoldDogan}
  P.~B.~Arnold and C.~Dogan,
  Phys.\ Rev.\  {\bf D78}, 065008 (2008)
  [arXiv:0804.3359 [hep-ph]].

\bibitem{ArnoldXiao}
  P.~B.~Arnold and W.~Xiao,
  Phys.\ Rev.\  {\bf D78}, 125008 (2008)
  [arXiv:0810.1026 [hep-ph]].

 

  

    
    

  
  
  

    

  

  


  

\end{thebibliography}
%
%

\end{document}